\documentclass[12pt,oneside]{article}
\usepackage{epsfig, cite}
\usepackage{color}
\usepackage{ifthen}
\usepackage{graphicx}

\def\({\left(}
\def\){\right)}
\def\[{\left[}
\def\]{\right]}

\def\e{\begin{equation}}
\def\q{\end{equation}}
\def\m{\begin{eqnarray}}
\def\n{\end{eqnarray}}

\begin{document}
\thispagestyle{empty} \setcounter{page}{0}

\vspace{2cm}

\begin{center}
{\huge Simplified Chain Inflation}

\vspace{1.4cm}

Qing-Guo Huang

\vspace{.2cm}

{\em School of physics, Korea Institute for
Advanced Study,} \\
{\em 207-43, Cheongryangri-Dong,
Dongdaemun-Gu, } \\
{\em Seoul 130-722, Korea}\\
\end{center}

\vspace{-.1cm}

\centerline{{\tt huangqg@kias.re.kr}} \vspace{1cm}
\centerline{ABSTRACT}
\begin{quote}
\vspace{.5cm}

We propose a simplified chain inflation model and calculate the
primordial power spectra of the scalar and tensor fluctuations. The
spectral index and the tensor-scalar ratio are respectively 0.972
and 0.089 which are consistent with present cosmological
observations.

\end{quote}
\baselineskip18pt

\noindent

\vspace{5mm}

\newpage

Inflation models proposed by Guth in \cite{Guth:1980zm} not only
explain the large-scale homogeneity and isotropy of the universe,
but also provide a natural mechanism to generate the observed
magnitude of inhomogeneity (see \cite{Mukhanov:1990me,Lyth:1998xn}
etc. for a review). During the period of inflation, quantum
fluctuations are generated within the Hubble horizon, which then
stretch outside the horizon to become classical. In the subsequent
deceleration phase after inflation these frozen fluctuations
re-enter the horizon, and seed the matter and radiation density
fluctuations observed in the universe.

In the last two decades many inflation models were proposed.
However, none of them can be naturally realized in a fundamental
theory. Gravity governs the background evolution during inflation
and quantum effects provide a natural mechanism to generate the
primordial perturbations to seed the fluctuations for forming the
large scale structure of the universe and the anisotropies in cosmic
microwave background radiation which can be measured precisely. A
quantum theory of gravity is needed before we understand the full
theory in the early universe. Nowadays string theory, as we know, is
the only self-consistent quantum theory of gravity. String theory
can only live in ten-dimensional spacetime. We need to compact
ten-dimensional string theory to four dimensions. Many free
parameters appear after compactifications, even though there is no
free dimensionless parameter in string theory in ten dimensions.
Recent developments for the flux compactifications
\cite{Giddings:2001yu,Kachru:2003aw} suggest that a huge number of
meta-stable string vacua emerge in string theory. The whole space of
such string vacua is called string landscape.

Cosmological observations implies that the density perturbation is
roughly ${\delta\rho / \rho}\sim 10^{-5}$ in our universe. A flat
potential of the inflaton is called for in inflation model. On the
other hand, a flat potential naturally brings a large enough number
of e-folds during inflation to solve the problems of hot big bang
model. Usually we can expect that one or more small dimensionless
parameters in the potential of inflaton is related to the small
density perturbation. On question we should ask is why there is such
a small parameter in a fundamental theory. Since the number of the
meta-stable vacua in string landscape is so huge, it offers an
opportunity to explain this small number. Actually many inflation
models in string theory, such as
\cite{Kachru:2003sx,Blanco-Pillado:2004ns,Blanco-Pillado:2006he,Allahverdi:2007wh,Otha}
and so on, have been proposed in the last few years. But what is the
distinguishing phenomenon for string landscape is still an open
question. The authors in \cite{Freivogel:2005vv} suggested that the
string landscape has a model independent prediction which says that
our universe should be spatially open.

Even though the shape of string landscape has not been figured out,
heuristically we expect that the meta-stable vacuum with large
vacuum energy has a very short life time. In \cite{HenryTye:2006tg},
Tye suggested that our universe would decay rapidly from a site with
large vacuum energy to a long-lived meta-stable vacuum with a small
positive cosmological constant through the resonance tunneling. Thus
it is possible to dynamically solve today's dark energy problem.
However, inflation is still needed in the early universe.
Fortunately, chain inflation
\cite{Freese:2004vs,Freese:2005kt,Freese:2006fk} becomes generic in
this scenario. In this model, the universe tunneled rapidly through
a series of meta-stable vacua with different vacuum energies. After
many tunneling events, more than 60 e-folds are obtained and the
problems in hot big bang are solved. In \cite{Feldstein:2006hm}, the
authors use numerical method to calculate the amplitude of the
primordial power spectra. However, their simulating results implies
a quite large tensor-scalar ratio which is not consistent with
cosmological observations. \footnote{In \cite{Feldstein:2006hm}, the
amplitude of the scalar power spectrum is given by $\Delta_{\cal
R}^2\sim \beta/\alpha^2$. On the other hand, the power spectrum for
the tensor modes is roughly the same as that for the perturbation of
the scalar field. The amplitude of the tensor perturbations in
\cite{Feldstein:2006hm} is $\Delta_T^2\sim \beta$. Thus the
tensor-scalar ratio takes the form $r=\Delta_T^2/\Delta_{\cal
R}^2\sim \alpha^2$. However, their simulating results shows that
$\alpha$ is much larger than 1, which contradicts to WMAP3
\cite{Spergel:2006hy} where $r\leq 0.65$ ($95\%$ CL).}

In this short note, we propose a simplified chain model and suggest
a new method to compute the density perturbations, spectral index
and tensor-scalar ratio. Our results perform a nice fit to the data.

In chain inflation model, the universe begins in a meta-stable
vacuum with a large positive vacuum energy. The system could tunnel
to the lower minima along a variety of possible directions in the
string landscape. However, many of the neighboring minima is
inaccessible, since the tunneling probabilities into them are too
small. In \cite{Coleman:1977py,Callan:1977pt}, the authors
investigated a classical field theory in which there are two
homogeneous equilibrium states with different energy densities. In
the quantum version of the theory, the state of higher energy
density becomes unstable through barrier penetration. To be simple,
we only quantified the tunneling probability by modeling a single
tunneling event for a scalar field theory with potential \e
V(\phi)={1 \over 4}\lambda (\phi^2-a^2)^2+{\sigma \over 2a}(\phi-a),
\q where $\sigma$ is the vacuum energy difference between two
neighboring meta-stable vacua. The tunneling probability is roughly
given by \e \Gamma\sim \exp \(-{27\pi^2\over 2}{\kappa^4\over
\sigma^3} \),\q where $\kappa$ is the tension of the brane
interpolating between the two vacua. In
\cite{Guth:1982pn,Turner:1992tz}, Guth et al. showed that the
probability of a point remaining in a false de Sitter vacuum is
roughly given by \e p(t)\sim e^{-{4\pi\over 3}\beta Ht}, \q here the
dimensionless parameter $\beta$ is \e \beta={\Gamma \over H^4}, \q
and $H$ is the Hubble parameter in this vacuum. Thus the lifetime of
the field in this meta-stable vacuum is estimated as \e \tau\simeq
{3\over 4\pi H\beta}. \q A lower bound on the dimensionless
parameter $\beta$ is obtained \e \beta\geq {9\over 4\pi}\q in order
for percolation and thermalization to be achieved. In this paper, we
focus on the case with the lifetime of the meta-stable vacua much
short than Hubble time $H^{-1}$, or equivalently $\beta\gg {3\over
4\pi}$.

In the simplified chain inflation we assume $\sigma$ is always a
constant. Since inflation responsible for the observed modes of the
density perturbations only lasts a few e-folds, the Hubble parameter
$H$ can be taken as a constant during this period. Naively the
vacuum energy goes like \e \rho_V\simeq \rho_{V,i}-{\sigma\over
\tau}t, \q where $\rho_{V,i}$ is the initial vacuum energy. The
vacuum energy drops $\sigma/ (H\tau)$ per Hubble time $H^{-1}$. Here
we also assume $\sigma$ is much smaller than $\rho_V$ and then the
series of tunneling events can be regarded as a continuous evolution
of a scalar field. The Hubble parameter is governed by Friedmann
equation \e H^2={\rho_V\over 3M_p^2}={\rho_{V,i}-{\sigma\over
\tau}t\over 3M_p^2},\label{hb} \q where $M_p$ is the reduced Planck
scale. The number of e-folds before the end of inflation is related
to the energy density $\rho_V$ by \e
N_e=\int_{t}^{t_{end}}H(t')dt'\simeq {2\over 3\sqrt{3}}{\tau \over
M_p\sigma}\rho_V^{3/2}, \label{nec}\q or equivalently, \m
\rho_V&\simeq& \({3\sqrt{3}\over 2}{M_p\sigma\over
\tau}N_e\)^{2\over 3}, \\ H^2&\simeq& \({\sigma/\tau\over
2M_p^2}N_e\)^{2\over 3}. \n According to eq. (\ref{hb}), the
``slow-roll'' parameter $\epsilon$ is given by \e \epsilon\equiv
-{\dot H\over H^2}={\sigma/\tau\over 6M_p^2H^3}={1\over 3N_e}.
\label{epl}\q  The meaning of $\sigma/\tau$ is nothing but the
change of the vacuum energy per time unit. Generically the tunneling
from one site to another site with a lower cosmological constant is
accompanied by some radiation due to the bubble percolation,
nucleation and collision. During inflation radiation is inflated
away. Suppose the change of the vacuum energy at the last step is
large, which implies that the change of the vacuum energy per time
unit is quite large and then the ``slow-roll" condition $\epsilon\ll
1$ is broken down. On the other hand, the loss of the vacuum energy
is transferred to radiation and then the energy density is dominated
by radiation. Now chain inflation ended and reheating happened.

Now the amplitude of the primordial power spectra for the scalar and
tensor perturbations can be expressed as respectively \m
\Delta_{\cal R}^2&=&{H^2/M_p^2\over 8\pi^2\epsilon}={3\over
2^{11/3}\pi^2}\({\sigma/\tau\over M_p^5}\)^{2\over 3}N_e^{5\over 3}, \\
\Delta_T^2&=&{H^2/M_p^2\over \pi^2/2}={2^{1/3}\over
\pi^2}\({\sigma/\tau\over M_p^5}\)^{2\over 3}N_e^{2\over 3}. \n The
spectral index and the tensor-scalar ratio are given by \m
n_s&=&1+{d \ln \Delta_{\cal R}^2\over d \ln k}\simeq 1-{d \ln
\Delta_{\cal R}^2\over d N_e}=1-{5\over 3N_e}, \\
r&=&{\Delta_T^2\over \Delta_{\cal R}^2}={16\over 3N_e}. \n For
$N_e=60$, the spectral index and the tensor ratio are respectively
$n_s=0.972$ and $r=0.089$. The results of WMAP three-year data are
presented in \cite{Spergel:2006hy}. To properly compare our results
with WMAP results, we show our results in Fig. 1.
\begin{figure}[h]
\begin{center}
\epsfxsize=0.7\columnwidth \epsfbox{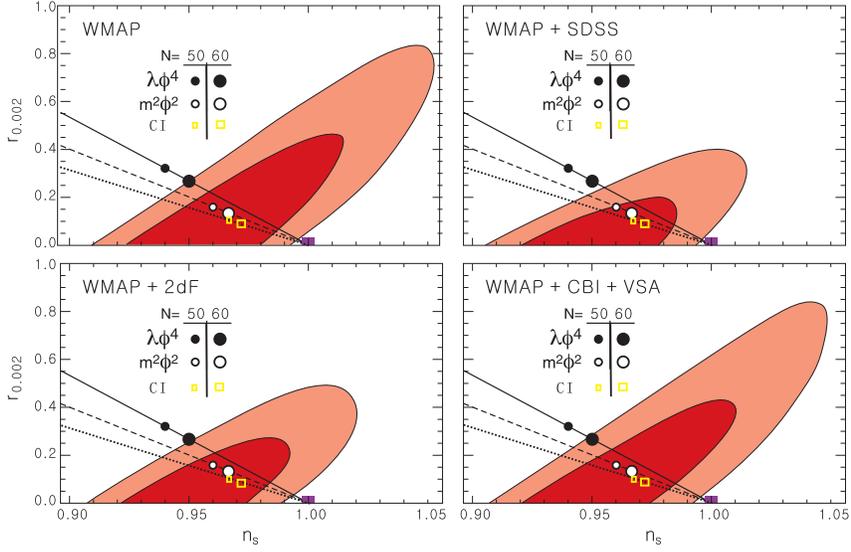}
\end{center}
\caption{``CI" denotes chain inflation.}
\end{figure}
Our results are nicely compatible with WMAP at $95\%$ confidence
level.

Using the WMAP normalization $\Delta_{\cal
R}^2=19.9_{-1.8}^{+1.3}\times 10^{-10}$, we find \e
{\sigma/\tau\over M_p^5}=8.6\times 10^{-16}. \q The Hubble parameter
and the energy density at $N_e=60$ are respectively $H\simeq 3\times
10^{-5} M_p$ and $\rho_V\simeq 2.7\times 10^{-9} M_p^4$. Requiring
$\sigma\ll \rho_V$ yields \e {\tau \over t_p}\ll 3\times 10^6, \quad
\hbox{or} \quad {\tau \over H^{-1}}\ll 90, \q where $t_p=M_p^{-1}$
is the Planck time. Our requirement that the lifetime of the
meta-stable vacua is much shorter than Hubble time $H^{-1}$ is
reasonable.

To summarize, a simplified chain inflation model is proposed. In
this model, the vacuum energy drops only a little per step and the
series of tunneling just looks like the continuous evolution. We
also estimate the density perturbations and we find our model can
fit the present data very well.

In the simplified chain inflation, we assume the vacuum energy drops
the same energy density $\sigma$ in each tunneling event. However we
do not expect it really happens in string landscape where the
situation becomes much more complicated. It is even possible that in
some steps along the chain the system tunnels to the vacua with
slightly higher energy density. But we can take our
semi-quantitative calculation as an averaged effect. Our results
only depend on $\sigma/\tau$. Naively we suggest that $\sigma/\tau$
should be replaced by the average value $\langle \sigma/\tau
\rangle$ in string landscape. Generically there are vast number of
chains, but only a few of them can satisfy what we have observed.
The amplitude of the scalar power spectra depends on the detail of
the chain. But the spectral index and the tensor-scalar ratio can be
taken as the quite insensitive predictions of string landscape.

\vspace{.5cm}

\noindent {\bf Acknowledgments}

We would like to thank H. Tye for useful comments. We also
acknowledge the use of one figure from WMAP collaboration.

\end{document}